\newcommand{\dint}{{\rm d}}
    \documentclass[%
    reprint,
    amsmath,amssymb,
    aps,
]{revtex4-1}
    \usepackage{amsmath}
    \usepackage{cleveref}
    \usepackage{amsfonts}
    \usepackage{graphicx}
    \usepackage{dcolumn}
    \usepackage{bm}

    \usepackage{titlesec}
  \usepackage{easyReview}
    \titlespacing{\subsubsection}{2pt}{\parskip}{-\parskip}
    
\begin{document}

    	\title{ Geometric scaling in leading neutron events at HERA}

    	\author{Arjun Kumar}
    	\email{arjun.kumar@physics.iitd.ac.in}

    	\affiliation{%
    		Department of Physics, Indian Institute of Technology Delhi, Hauz Khas, New Delhi 110016, India\\
    	}

    	\date{\today}

    	\begin{abstract}
	
	
	This analysis provides new fits of the GBW model and the impact parameter-dependent saturation model (bSat or IP-Sat) to the leading neutron structure function HERA data in one pion exchange approximation. Both parametrizations of the dipole cross section provide good descriptions of the considered data. It is shown here for the first time that the experimental leading neutron production HERA data exhibits geometric scaling, which in this context means that the total $\gamma^* \pi^*$ cross section is a function of only one dimensionless variable $\tau = Q^2/Q_s^2(x)$. The geometric scaling region extends over a broad range of $Q^2$ and can be attributed to the presence of a saturation boundary which manifests at $Q^2\geq Q^2_s$. The scaling behaviour in leading neutron events is profoundly similar to what has been observed for the inclusive DIS events.
    	\end{abstract}

    	\maketitle

    	\section{\label{introduction}INTRODUCTION}
    	Over the last few decades, deep inelastic scattering (DIS) experiments have been essential in understanding the structure of the proton. The H1 and ZEUS experiments at HERA measured the proton structure function at an unprecedented precision to date. Mostly, this data is recorded in bins of the Bjorken variable $x$ and the photon virtuality $Q^2$ and available in a wide kinematic region reaching very small values of $x \sim 10^{-7}$. Though $x$ and $Q^2$ are independent variables, the total cross section $\sigma^{\gamma^*p}_{\text{tot}}$ extracted from this inclusive data shows a striking feature that it depends only one dimensionless variable $\tau = Q^2R_s^2(x)$ at low $x$.  This was first observed by Sta\'sto, Golec-Biernat and Kwieci\'nski in \cite{PhysRevLett.86.596} and is commonly known as "geometric scaling". This scaling behavior has a natural explanation in the dipole models with saturation for photon virtualities smaller than the saturation scale ($Q_s^2$) but for $Q^2>Q_s^2$ this is not associated with saturation physics, rather this regularity exists in solution of evolution equations as demonstrated by Iancu, Itakura and Mclerran for BFKL equations in \cite{IANCU2002327} and in DGLAP equations with initial condition provided along the critical line $Q^2=Q_s^2(x)$  as discussed in detail in \cite{PhysRevD.66.014013}. Further in \cite{Caola:2008xr}, the  authors argued that the geometric scaling can be explained with generic boundary conditions for the standard DGLAP evolution. In general, the  geometric scaling is expected to hold for ln$Q^2/Q^2_s(x)<<$ ln$Q^2_s(x)/\Lambda_{QCD}^2$, usually referred to as the extended geometric scaling regime.    More detailed investigations of scaling behavior in inclusive DIS events are provided in \cite{Gelis:2006bs,Avsar:2007ht,Beuf:2008mf,Caola:2010cy,Praszalowicz:2012zh}. In addition, the diffractive DIS data also exhibits similar scaling behavior \cite{Marquet:2006jb}.

    	In some of the DIS events, baryons carrying large fraction of the proton's longitudinal momentum ($x_L >0.3$) are produced in the far forward direction, commonly known as the leading baryons. These kind of events have been observed at HERA with leading neutrons, protons and photons \cite{H1:2010hym,H12014data, ZEUS:2002gig,collaboration_2009}. Recently, the dipole framework has been extended to study the leading neutron events employing the one pion exchange (OPE) approximation \cite{Goncalves:2015mbf,Carvalho:2015eia,Amaral:2018xsm,Carvalho:2020myz,PhysRevD.105.114045,BISHARI1972510}. In the dipole picture, the virtual photon emitted from the incoming electron splits into quark-antiquark pair forming a color dipole which then interacts with the target via strong interaction. In the case of leading neutrons, the color dipole interacts with the pion cloud of the proton, and the forward neutron comes from the proton as it splits into a neutron and a positive pion. 
    	
    	In our recent study \cite{PhysRevD.105.114045}, we showed using the saturated and non-saturated impact parameter dependent dipole models that the leading neutron  data is insensitive to non-linear effects. This could be understood as the Bjorken $x$ value probed in this semi-inclusive measurement is considerably larger than the Bjorken $x$ in proton DIS events, where the latter has exhibited no clear signal for saturation. The next crucial step in this direction is to check whether the leading neutron events exhibit geometric scaling as observed in inclusive DIS events. This has not been tested thus far and  is an important step from a phenomenological point of view. This paper aims at investigating the leading neutron events to find whether or not this regularity is observed in the experimental data.  Here two well known parametrisation of the dipole models with saturation are used; the original Golec-Biernat and
    	W\"usthoff (GBW) model \cite{PhysRevD.59.014017} and the bSat (IP-Sat) model \cite{Kowalski:2003hm,Kowalski:2006hc} which has an explicit DGLAP evolution. To obtain the saturation scale,  new fits of the leading neutron structure function data  are performed with both the models employing OPE. 
    	    	\begin{figure*}
    		\includegraphics[width=0.35\linewidth]{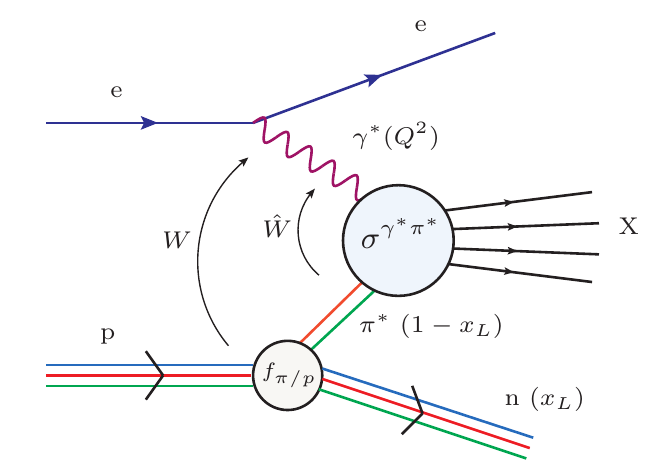}\hskip2cm
    		\includegraphics[width=0.35\linewidth]{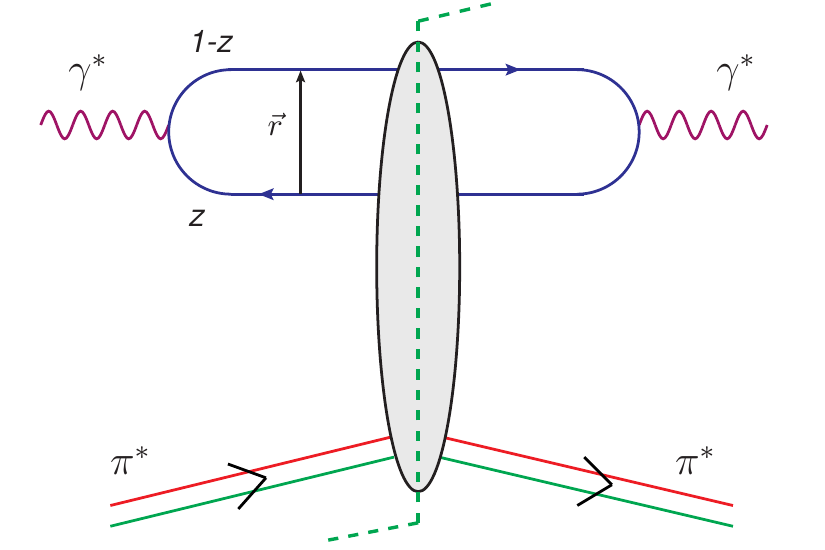}
    		\caption{Leading neutron production in one-pion exchange approximation in ep collisions (a) and $\gamma^*\pi^*$ scattering cross section in dipole model (b)}
    		\label{dipolepion}
    	\end{figure*}
    	
    	The rest of the paper is organised as follows. In the next section, a brief outline of the leading neutron production in GBW and bSat models is given and the fitting procedure is discussed. In section 3, the fit results, the extracted saturation scale and the geometric scaling results are presented. In section 4,  we summarise and discuss the main conclusions of our study.

\section{LEADING NEUTRON PRODUCTION IN  THE DIPOLE MODELS}
The dipole framework is formulated in the target rest frame where the incoming electron emits a photon which splits into a quark antiquark  pair forming a color dipole which subsequently interacts with the target strongly. For leading neutron production, the dipole probes the pion cloud of the proton where the virtual pion comes from the proton as it splits into a neutron and a pion as illustrated in Fig.~\ref{dipolepion}. In the one-pion exchange approximation, at high energies, the differential cross section for $\gamma^*p\rightarrow Xn$ can be written as \cite{Carvalho:2015eia}:
\begin{equation}
\frac{\dint^2 \sigma (W,Q^2,x_L,t)}{\dint x_L \dint t } = f_{\pi/p}(x_L,t) ~\sigma^{\gamma^* \pi^*}(\hat{W^2},Q^2)
\end{equation}
where $t$ is the four-momentum transfer squared at the proton vertex, $x_L$ is the proton’s longitudinal momentum fraction taken by the neutron, $W$ is the centre-of-mass energy for the photon-proton system, $\hat{W}$ is the centre-of-mass energy for the photon-pion system with $\hat{W}^2=(1-x_L)W^2$, $ f_{\pi/p}(x_L,t)$ is the flux of pions emitted by the proton and $\sigma^{\gamma^* \pi^*}$ is the cross section of $\gamma^*\pi^*$ interactions. The $t$ variable is related to $p_T$, the transverse momentum of the neutron, and $x_L$ as:
\begin{equation}
t \simeq -\frac{p_T^2}{x_L}-(1-x_L)\left(\frac{m_n^2}{x_L}-m_p^2\right)
\end{equation}
where $m_n$ and $m_p$ are the masses of neutron and proton respectively. The leading neutron structure function is given by \cite{H1:2010hym}:
\begin{eqnarray}
F_2^{LN}(W,Q^2,x_L)= \Gamma(x_L,Q^2) F_2^{\pi}(W,Q^2,x_L) 
\end{eqnarray}
Here $\Gamma(x_L,Q^2) = \int_{t_{min}}^{t_{max}} f_{\pi/p}(x_L,t)\dint t$ is the pion flux factor integrated over the \emph{t}-region of the measurement and $ F_2^{\pi}(W,Q^2,x_L) =  \frac{ Q^2}{4 \pi^2 \alpha_{\rm EM}}\sigma^{\gamma^*\pi^*}(\hat{W^2},Q^2) $ is the pion structure function.
\subsection{The pion flux}
The flux factor $ f_{\pi/p}(x_L,t)$  describes the splitting of a proton into a $\pi n$ system. 
Following \cite{PhysRevD.105.114045,Goncalves:2015mbf,Carvalho:2015eia, Amaral:2018xsm,Carvalho:2020myz}, the flux factor given by:
\begin{equation}
f_{\pi/p}(x_L,t)=\frac{1}{4 \pi}\frac{2 g^2_{p\pi p}}{4 \pi} \frac{|t|}{(m_\pi^2+|t|)^2}(1-x_L)^{1-2\alpha(t)}[F(x_L,t)]^2
\label{eq:flux}
\end{equation} 
where $m_\pi$ is the pion masss, $g^2_{p\pi p}/(4\pi)=14.4$ is the $\pi^0pp$ coupling. $F(x_L,t)$ is the form factor which accounts for the finite size of the vertex. The form factor is given as:
\begin{equation}
F(x_L,t) = \exp\bigg[-R^2\frac{|t|+m_\pi^2}{(1-x_L)}\bigg], \alpha(t) = 0
\end{equation}
where $R=0.6~$GeV$^{-1}$ has been determined from HERA data \cite{HOLTMANN1994363}.
%
%
%

	\begin{table*}
	
	\begin{tabular}{|c|c|c|c|c|c|c|}
		\hline 
		GBW&$\sigma_0$[mb]  & $\lambda$ & $x_0/10^{-4}$  &$R_q$& $\chi^2/N_{\text{dof}}$ \\ 
		\hline 
		Fit 1 &17.171 $\pm$  2.777 & 0.223 $\pm$ 0.018 & 0.036 $\pm$ 0.024&--& 63.26/48= 1.32\\ 
		\hline 
		Fit 2 &27.43  & 0.248  & 0.40 &0.438 $\pm$ 0.005& 64.52/50= 1.29\\ 
	
		\hline

	\end{tabular} \vskip0.4cm
	\begin{tabular}{|c|c|c|c|c|c|c|}
	\hline 
	bSat&$A_g$  & $\lambda_g$ & C  &$R_q$& $\chi^2/N_{\text{dof}}$ \\ 
	\hline 
	Fit 3 &1.208 $\pm$ 0.012&0.0600 $\pm$ 0.038&1.453 $\pm$ 0.024&~--~& 58.75/48 = 1.22\\ 
	
	\hline 
	Fit 4 & ~2.195~  &~ 0.0829 ~ & ~2.289~ &0.520 $\pm$0.006 & 66.19/50 = 1.32\\ 
	
	\hline 
	
\end{tabular} 
	\caption{Fit results of the GBW model and the bSat model to the leading neutron structure function HERA data for $\beta \leq 0.01$,   with $N_{\text{p}} = 51$ points for $x_{\text{L}}$$_{\text{min}}=0.6$. Quark masses are fixed and as given in  \cite{Mantysaari:2018nng}.}
	\label{table1}
\end{table*}

%
%

%
%

\subsection{ GBW model}
 Using the optical theorem, the total $\gamma^*\pi^*$ cross section is given by the imaginary part of the forward elastic $\gamma^*\pi^* \rightarrow \gamma^*\pi^*$ amplitude as following:
\begin{align}
\sigma_{L,T}^{\gamma^*\pi^*}(\beta,Q^2)=\int\dint^2\textbf{r}\int_{0}^{1}\frac{\dint z}{4 \pi} |\Psi^f_{L,T}(\textbf{r},z,Q^2)|^2 \sigma^{\pi}(\textbf{r},\beta)
\end{align}
where $\beta=\frac{Q^2+m_f^2}{(1-x_L)W^2+Q^2}$ is the scaled Bjorken variable for the photon-pion system. The photon wavefunctions are well known quantities calculated in \cite{Kowalski:2006hc} and $\sigma^{\pi}(\textbf{r},\beta)$ is the dipole-virtual pion cross section. The GBW model was proposed in \cite{PhysRevD.59.014017}, the first successful attempt to explain the inclusive HERA data in a saturation mechanism, the  dipole cross section in GBW model given by \cite{PhysRevD.59.014017}:
\begin{equation}
\sigma^{\pi}(r,\beta) =\sigma_0(1-e^{r^2 Q_s^2(\beta)/4})
\label{GBW}
\end{equation}
where the saturation scale $Q_s$ is defined as
\begin{equation}
Q_s^2(\beta)=Q_0^2(\beta/x_0)^{-\lambda}
\label{GBWQs}
\end{equation}
with $Q_0^2=1$ GeV$^2$. The above cross section has an  important property of "geometric scaling" \cite{PhysRevLett.86.596}, which means that it depends only on one dimensionless variable $rQ_s$ or $\tau$ defined as:
\begin{equation}
\tau = Q^2R_s^2(\beta) = \frac{Q^2}{Q_s^2(\beta)}
\label{scalevariable}
\end{equation}
 Here, there are three free parameters $\sigma_0, \lambda,  x_0$, now either these can be fitted  to the leading neutron structure function data or we could use the fit results of the inclusive proton data and make use of the assumption that the dipole-proton cross section and dipole-pion cross section are universal up to normalization at small $x$ \cite{Kopeliovich:2012fd, Kaidalov:2006cw,PhysRevD.105.114045}. This means
\begin{equation} 
\sigma^{\pi}(r,\beta) = R_q~ \sigma^{p}(r,\beta)
\label{GBWscaling}
\end{equation}
where $R_q$ is determined by fit to the leading neutron structure function data and the dipole-proton cross section is taken from usual DIS fit of the GBW model from \cite{Golec-Biernat:2017lfv}. Both these strategies  are used in this analysis and the fit results are provided in Table \ref{table1}.  For fit 1 of the GBW model, the parameters ($\sigma_0,~\lambda,$ and $x_0$) are fitted to the leading neutron structure function data, while for fit 2, these parameters are the same as determined from the usual inclusive DIS data and the assumption of hadronic universality at small $x$ between pions and protons (Eq.\eqref{GBWscaling}) is used and the only parameter is $R_q$, which is determined from the fit. Though this assumption works well as shown in \cite{PhysRevD.105.114045,NIKOLAEV2000157,PhysRevD.85.114025,DAlesio:1998uav}, it does not yield a physical saturation scale other than the dilute limit of this cross section. Hence, we will use the saturation scale extracted from fit 1 of the GBW model.

\begin{figure}
	\includegraphics[width=0.8\linewidth]{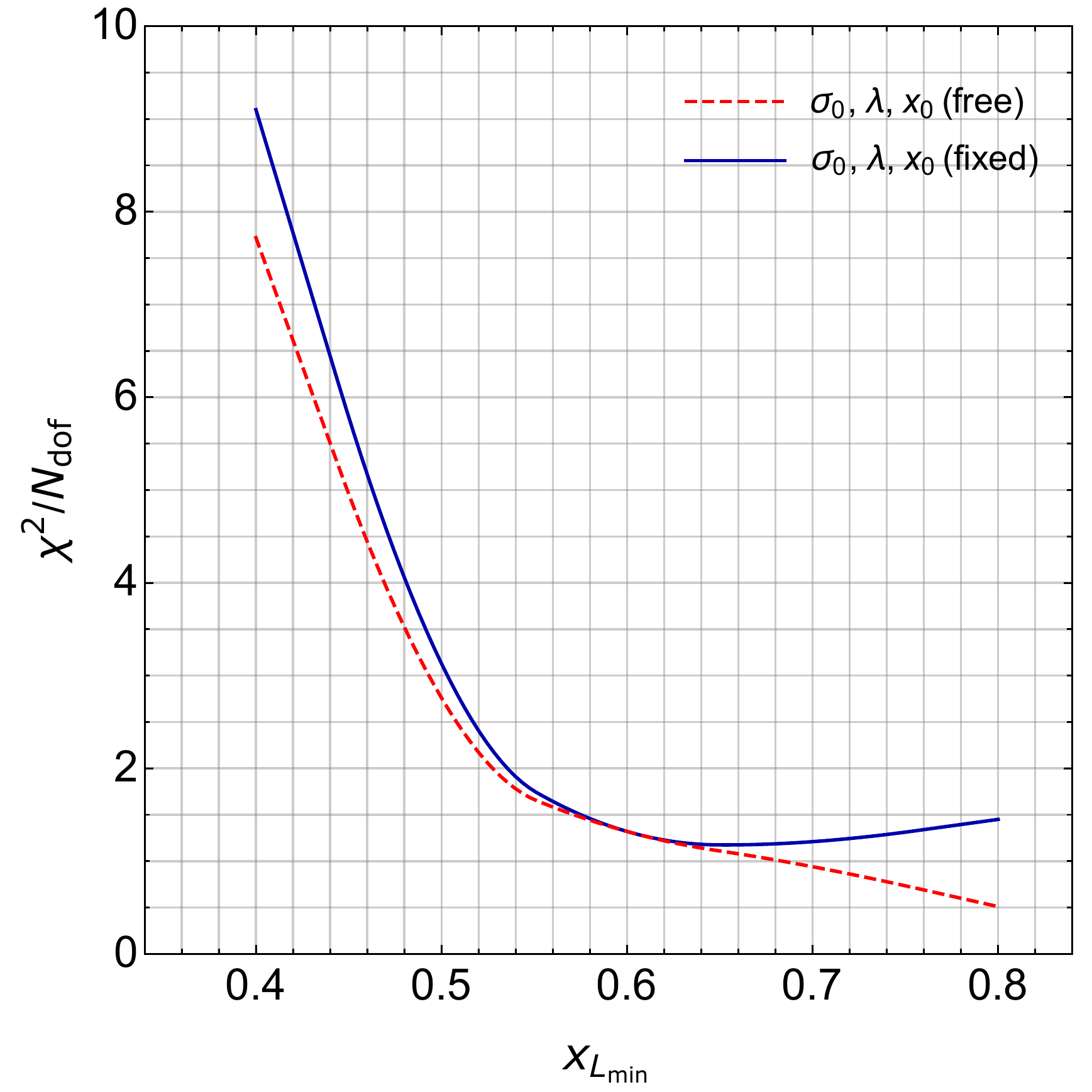}
	\caption{Sensitivity of the quality of Fit 1 from Table \ref{table1} to the choice of $x_{\text{L}}$$_{\text{min}}$ in the data. }
	\label{fit_details}
\end{figure}

\subsection{  bSat  model }
Using the optical theorem, the total $\gamma^*\pi^*$ cross section is given by:
\begin{align}
\sigma_{L,T}^{\gamma^*\pi^*}(\beta,Q^2) = &\int \dint^2\textbf{b}~\dint^2\textbf{r}\int_{0}^{1} \frac{\dint z}{4 \pi} |\Psi^f_{L,T}(\textbf{r},z,Q^2)|^2 \times \nonumber\\ &\frac{\dint\sigma^{(\pi)}_{q\bar{q}}}{\dint^2\textbf{b}}(\textbf{b},\textbf{r},\beta)
\end{align}
The bSat model contains DGLAP equation for evolution of gluon density for large scales and also has an explicit b-dependence. The dipole-pion cross section in the bSat model is given by \cite{Kowalski:2003hm,Kowalski:2006hc}:
\begin{eqnarray}
\frac{\dint\sigma^{(\pi)}_{q\bar{q}}}{\dint^2\textbf{b}}(\textbf{b},\textbf{r},\beta)=
2\big[1-\text{exp}\big(-F(\beta , r^2)T_p(\textbf{b})\big)\big]
\label{bsat}
\end{eqnarray}
with
\begin{eqnarray}
F(\beta , r^2)=\frac{\pi^2}{2N_C} r^2 \alpha_s(\mu^2) \beta g(\beta,\mu^2),
\end{eqnarray}

\begin{figure*}
	\includegraphics[width=0.35\linewidth]{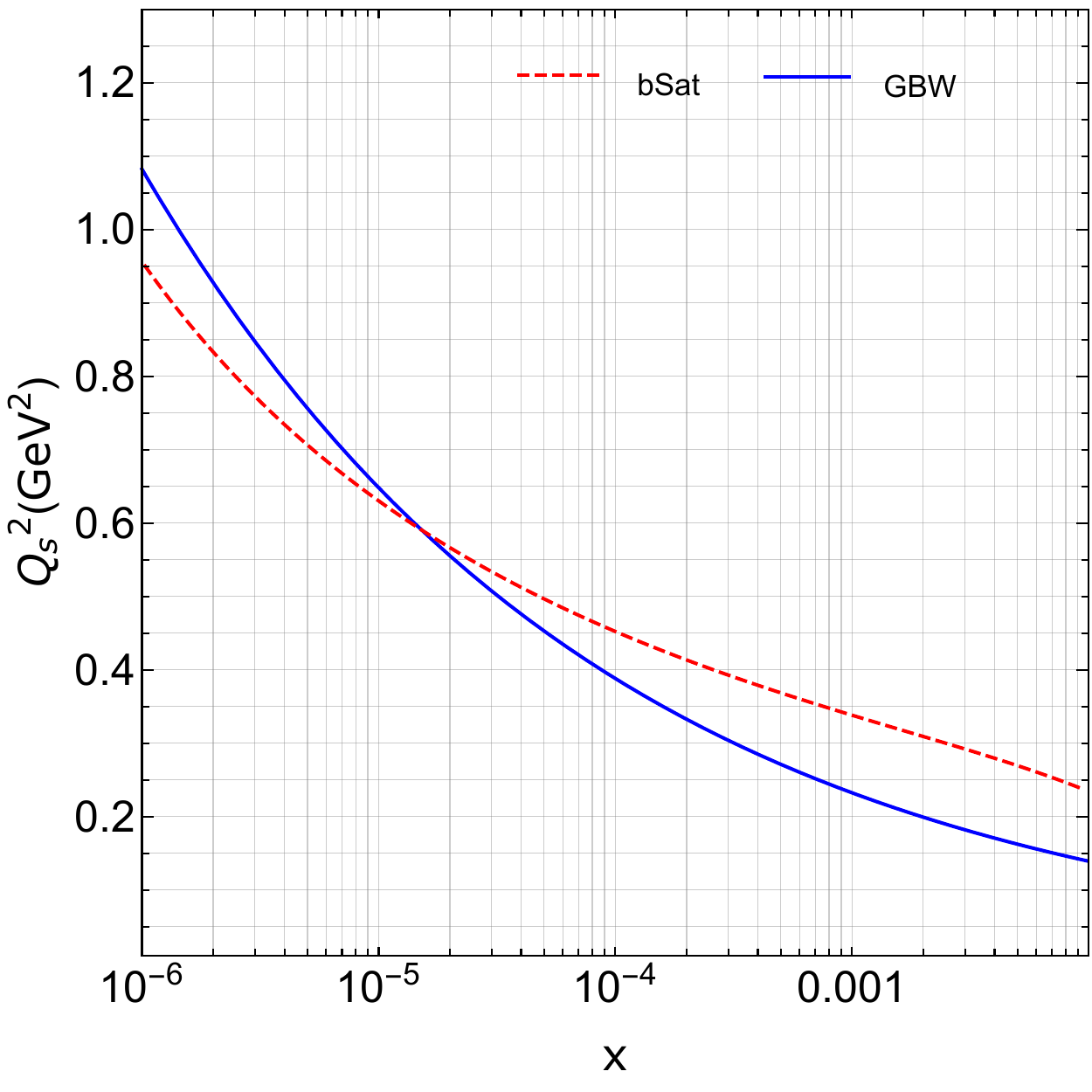}\hskip  1.2cm
	\includegraphics[width=0.35\linewidth]{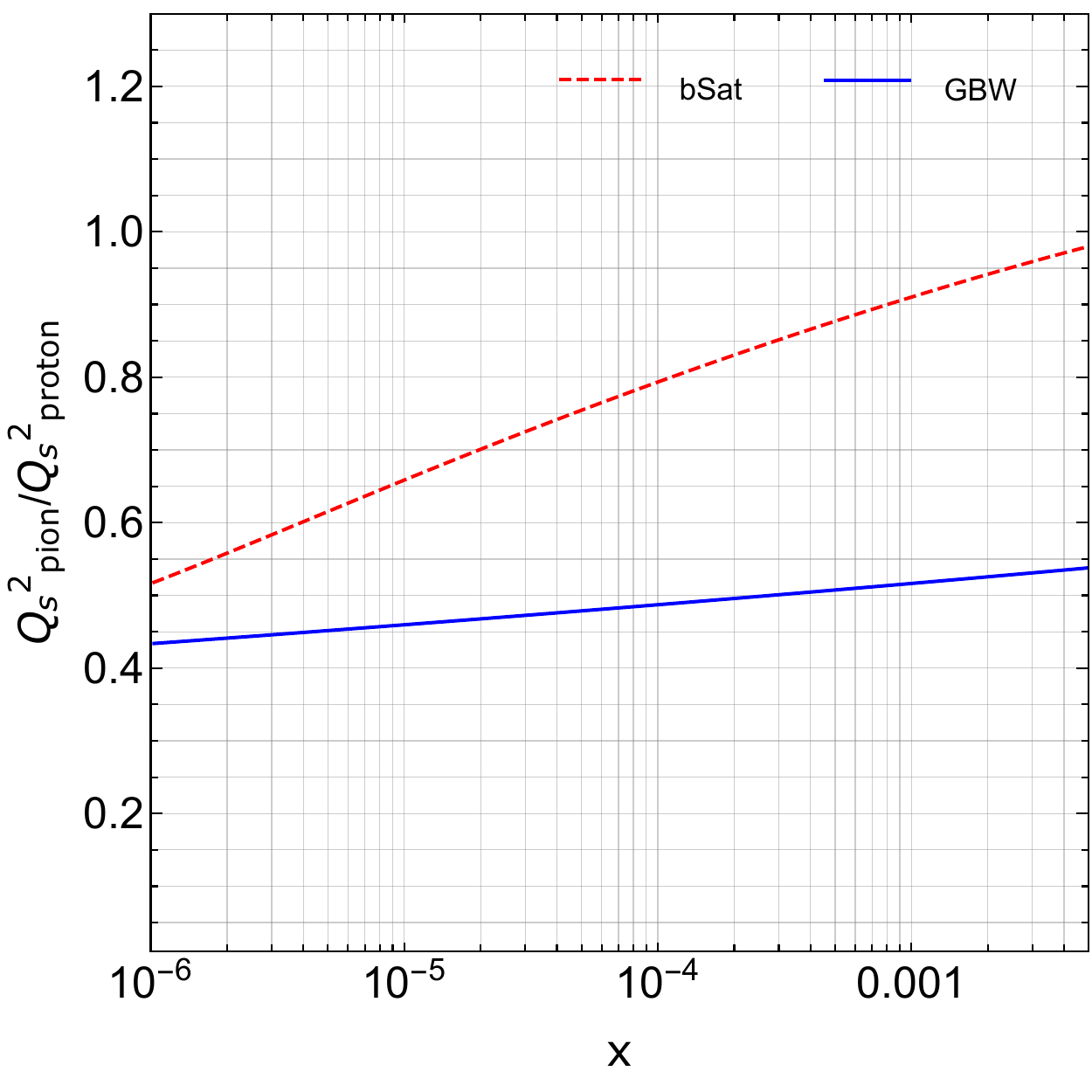}
	\caption{Saturation scale of pion as a function Bjorken $x$ (left) and ratio of saturation scale of pions to that of  protons (right) in GBW model  and bSat model (at b=0 fm)  at $x_{\text{L}}=0.6$ with the parameters from Fit 1 and Fit 3 in table \ref{table1} respectively. }
	\label{satscales}
\end{figure*}
The dipole cross section saturates for large dipole sizes $r$ and for large gluon densities.The scale at which the strong coupling $\alpha_s$ and gluon density is evaluated at is $\mu^2 = \mu_0^2 +\frac{C}{r^2}$ and the gluon density at the initial scale $\mu_0$ is parametrised as:
\begin{eqnarray*}
	\beta g(\beta,\mu_0^2)= A_g \beta^{-\lambda_g}(1-\beta)^{6}
\end{eqnarray*}
 The  transverse profile of the pion is assumed to be Gaussian with $T_{\text{p}}(\textbf{b}) = \frac{1}{2 \pi B_{\pi}}\exp\bigg(-\frac{\textbf{b}^2}{2B_\pi}\bigg)$ where $B_\pi $ is the  width of the pion. There is no available data which constrains this parameter. The width of pion is chosen  to be $B_\pi = 2$ GeV$^{-2}$, motivated from the Belle measurements \cite{Belle:2012wwz,Belle:2015oin} of hadron-pair production in a two-photon process $\gamma^* \gamma \rightarrow \pi^0 \pi^0$ where it was found to be $B_\pi=1.33-1.96~$GeV$^{-2}$ \cite{Kumano:2017lhr}. Also, H1 measurements \cite{H1:2015bxa} of the $\hat t$ spectrum for exclusive $\rho$ photo-production with leading neutrons in $ep$ scattering  suggests $B_\pi\approx 2.3$~GeV$^{-2}$. For more detailed discussion on probing the transverse width of pion experimentally see \cite{PhysRevD.105.114045}. Similar to the GBW model, we follow both methods; first performing an independent fit of the  gluon density parameters $A_g,~\lambda_g$ and $C$ to the leading neutron structure function data and secondly making use of the assumption:
\begin{equation} 
\frac{\dint\sigma^{(\pi)}_{q\bar{q} }}{\dint^2\textbf{b}}(\textbf{b},\textbf{r},\beta) = R_q~ \frac{\dint\sigma^{(p)}_{q\bar{q} }}{\dint^2\textbf{b}}(\textbf{b},\textbf{r},\beta)
\label{bSatscaling}
\end{equation}
where only $R_q$ is fitted to the leading neutron structure function data and the dipole-proton cross section is taken  from the fit of the bSat model to the usual DIS data from \cite{Mantysaari:2018nng}. Again, the saturation scale is extracted from the fit where  $A_g,~\lambda_g$ and $C$  are fitted to the leading neutron structure function data. The saturation scale in this case is given by \cite{Kowalski:2003hm}:
 \begin{equation}
  Q_S^2( \beta, b)\equiv 2/r_S^2
  \label{bSatQs}
 \end{equation}
where $r_S$ is defined by solving $1/2=F(\beta ,r_S^2)T_p(b)$ in the dipole amplitude.

\section{RESULTS}
 \begin{figure*}
	\includegraphics[width=0.35\linewidth]{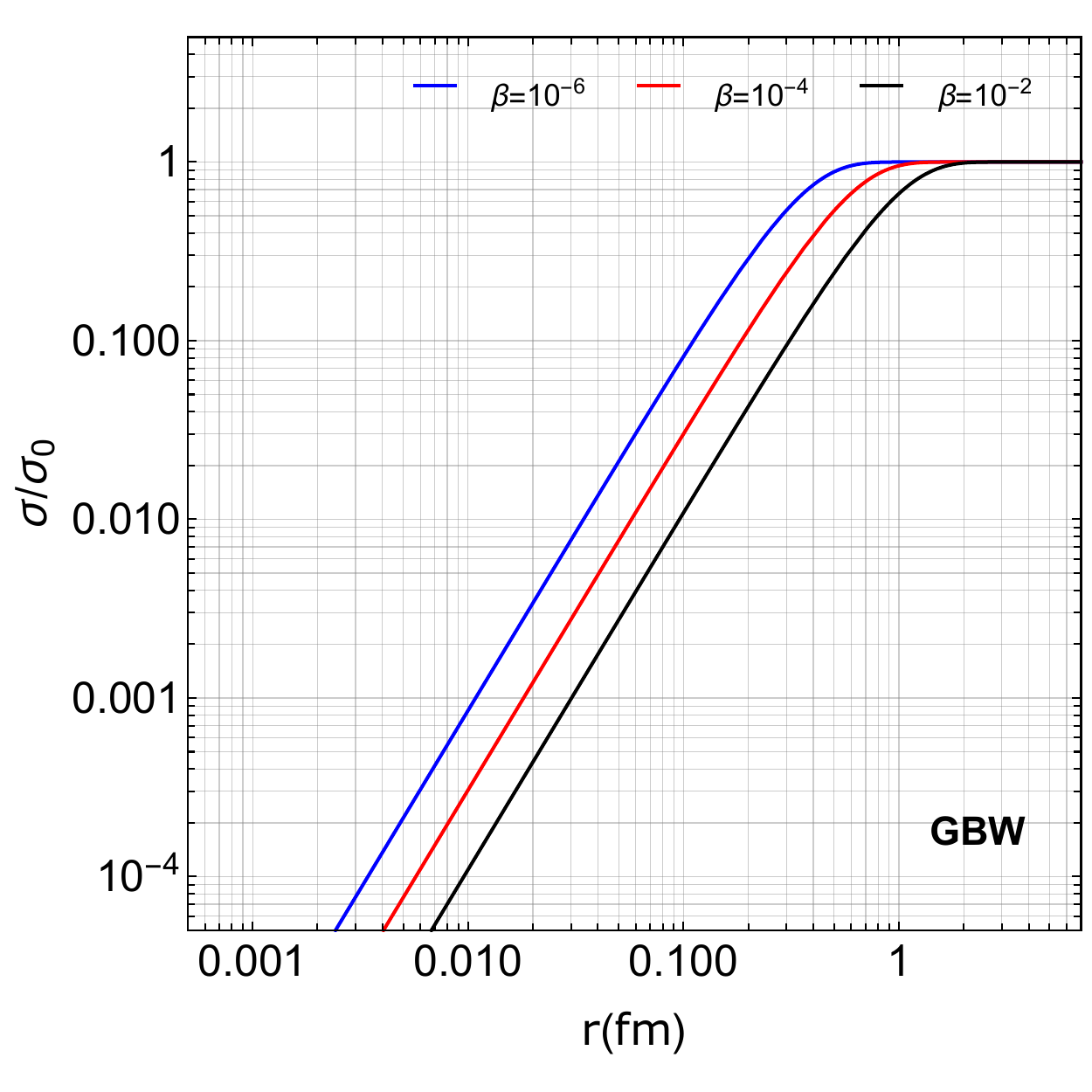}\hskip  1.2cm
	\includegraphics[width=0.35\linewidth]{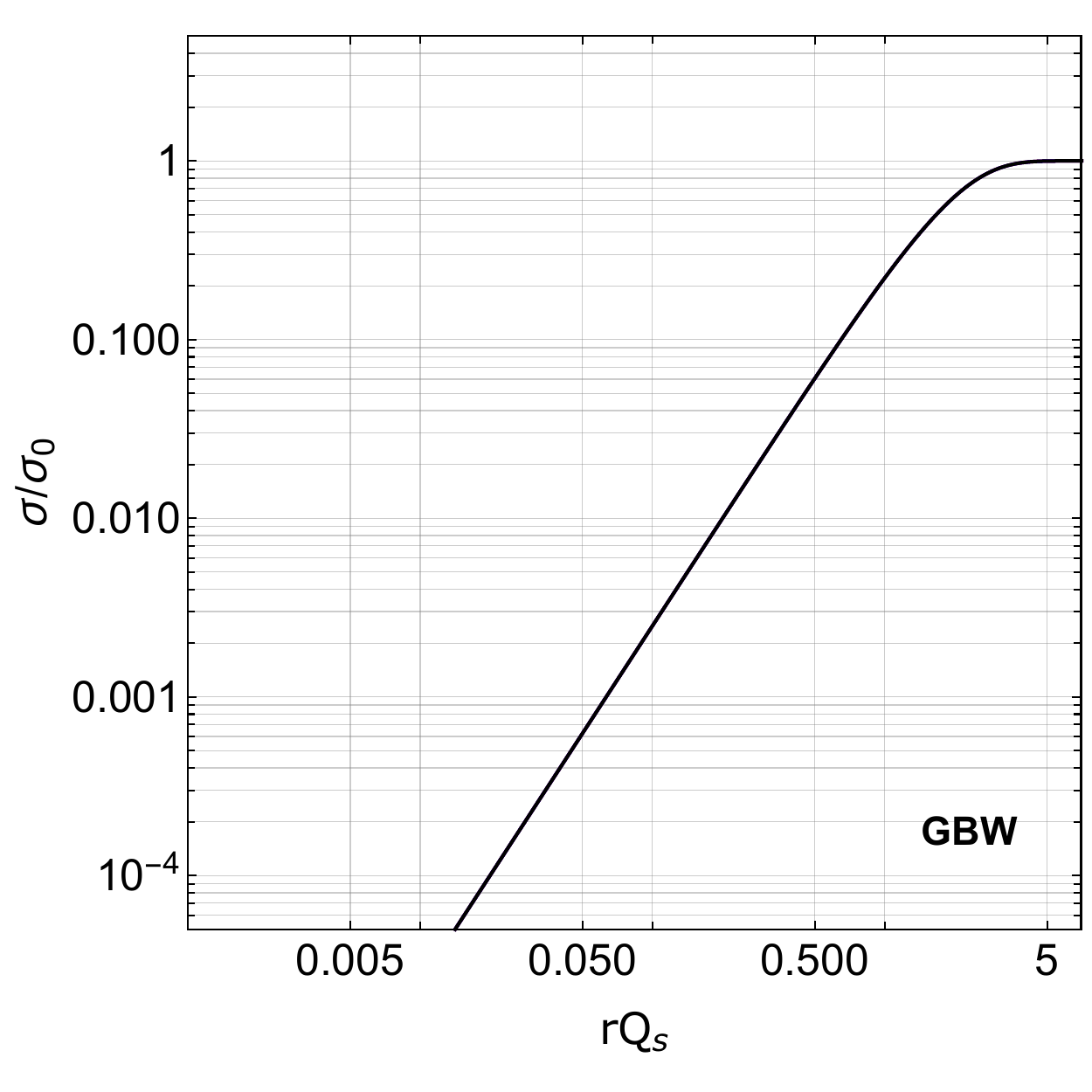}
	\caption{Normalised Dipole cross section (Eq.~\eqref{GBW}) with the parameters from Fit 1 in table \ref{table1} as a function $r$ (left)  and $rQ_s$ (right) for different values of $\beta$. As a result of geometric scaling, all the  curves from left plot merge into one curve in  right plot. }
	\label{sigmaGBW}
\end{figure*}
\begin{figure*}
	\includegraphics[width=0.35\linewidth]{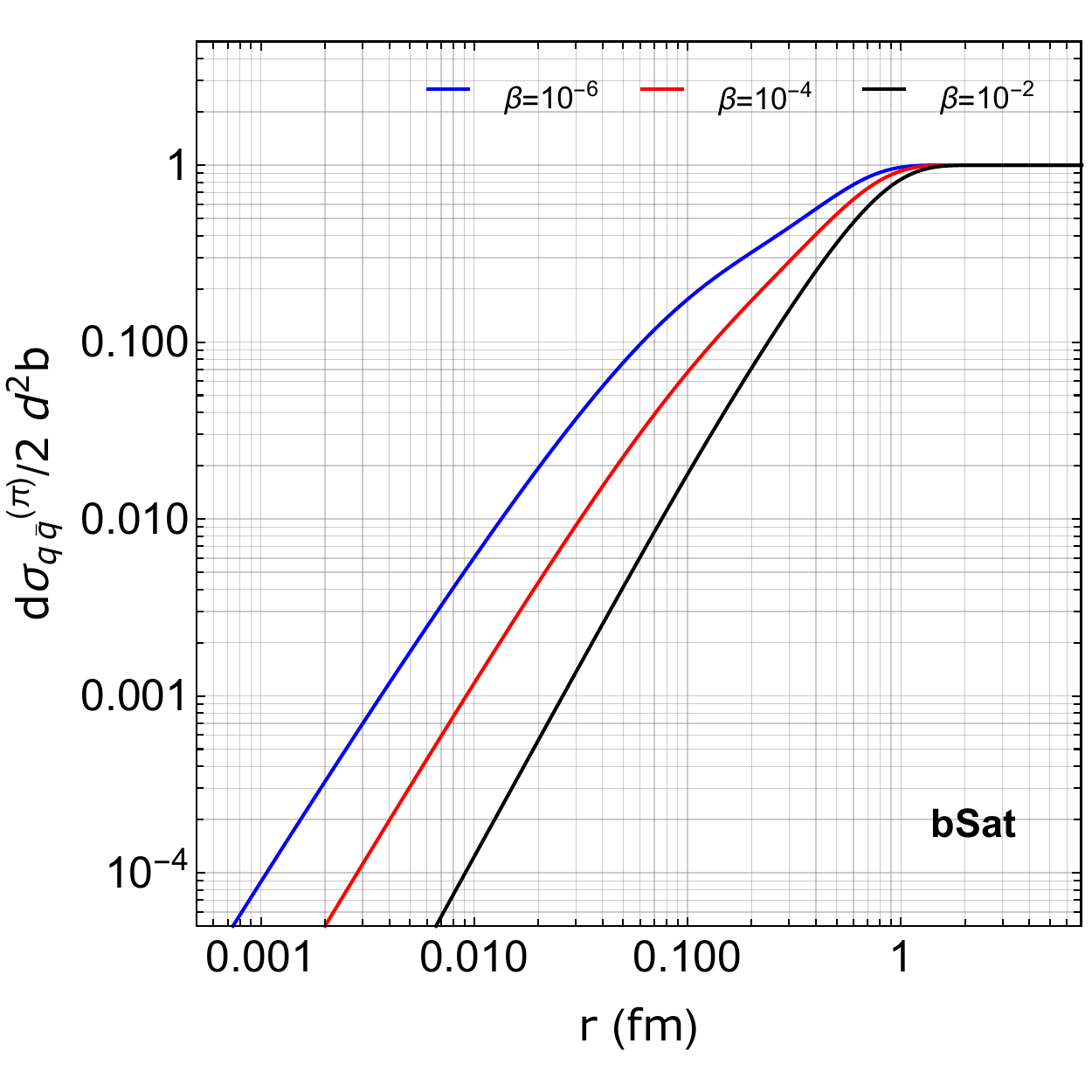}\hskip  1.2cm
	\includegraphics[width=0.35\linewidth]{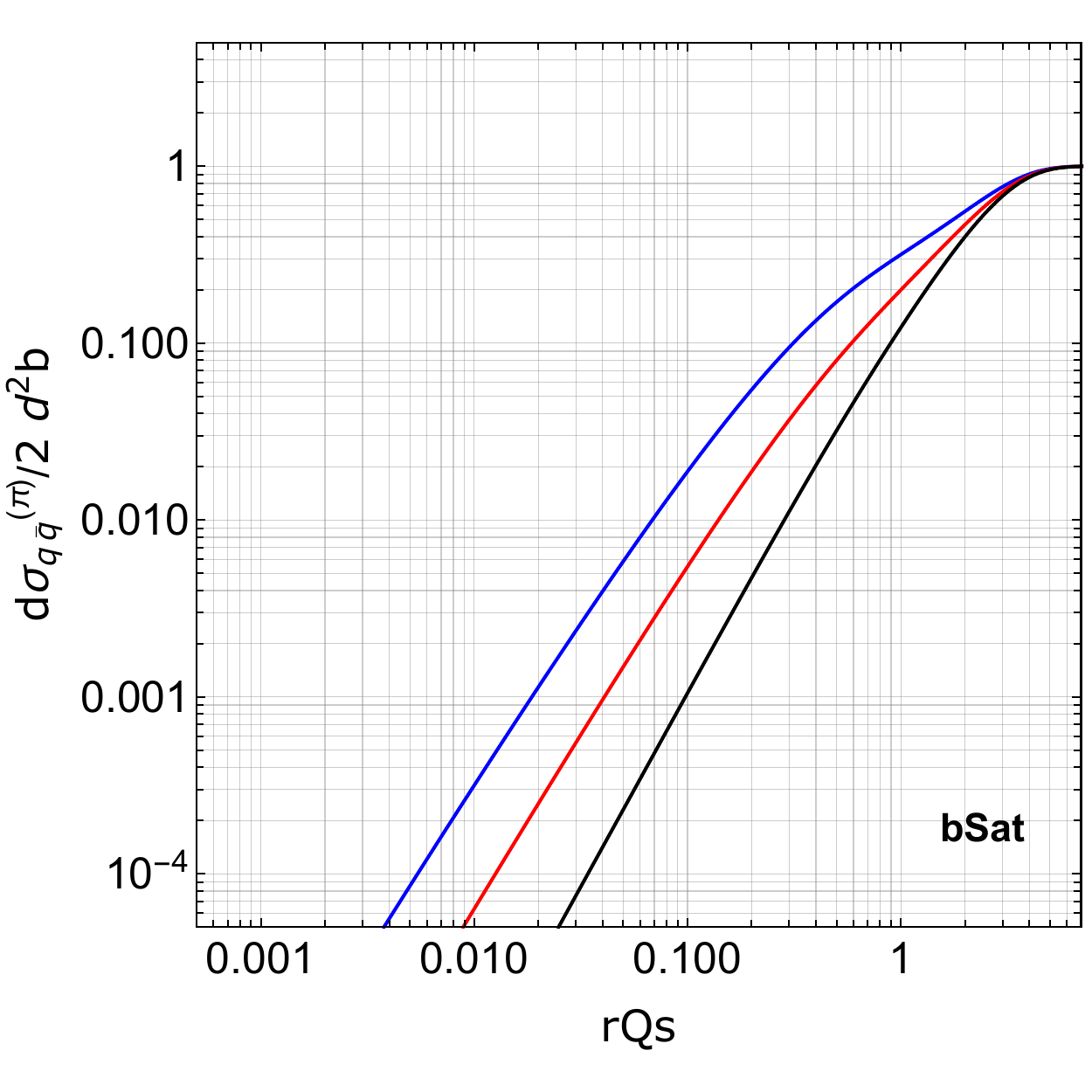}
	\caption{Dipole cross section (Eq.~\eqref{bsat}) with the parameters from Fit 3 in table \ref{table1} as a function $r$ (left)  and $rQ_s$ (right) for different values of $\beta$.  Due to the breaking of geometric scaling in bSat model, the curves from left plot do not merge into a  single curve in the right plot.}
	\label{sigmabSat}
\end{figure*}

\begin{figure*}
	\includegraphics[width=0.72\linewidth]{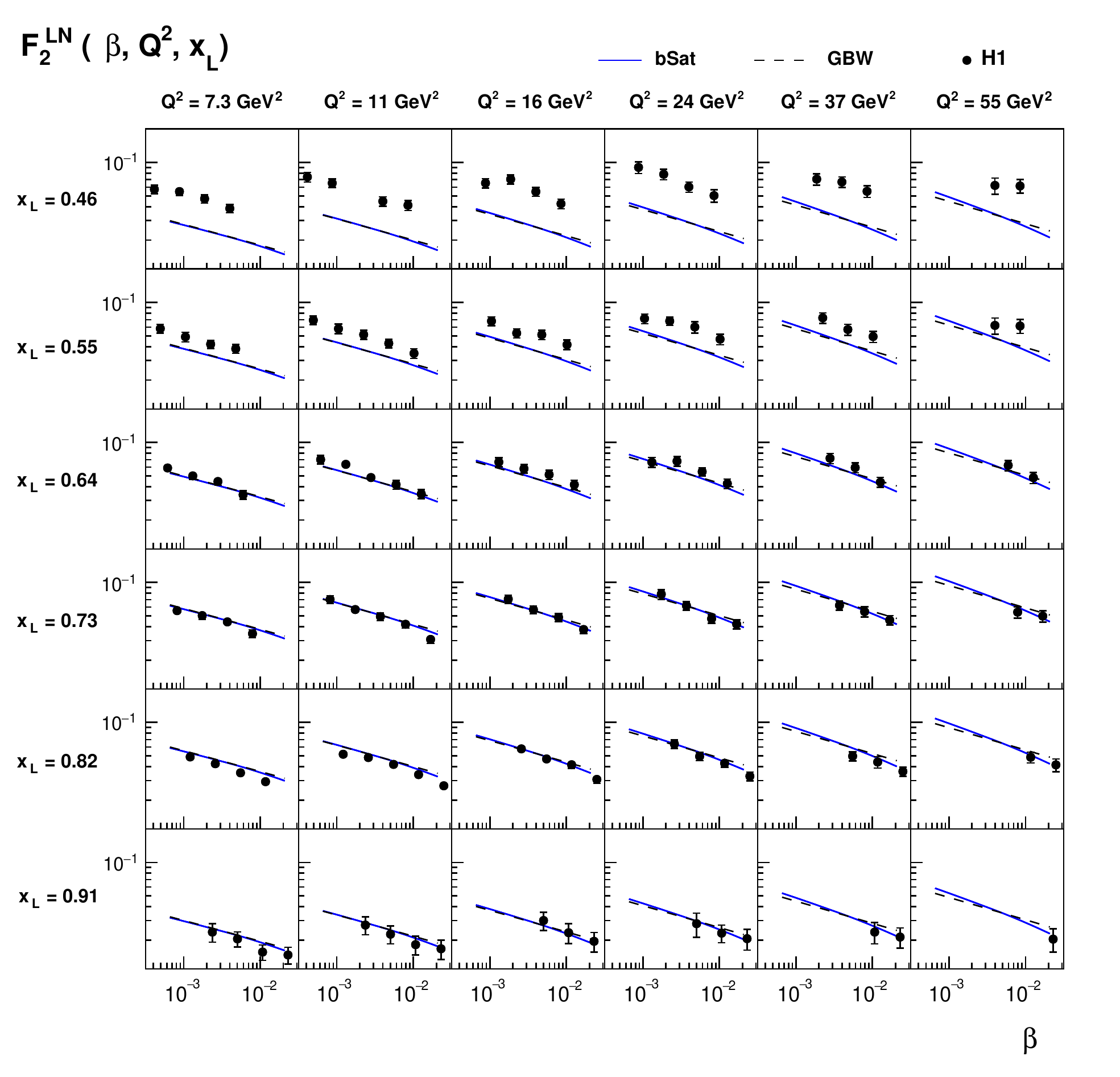}
	\caption{Comparison of the HERA data for leading neutron structure function $F_2^{LN} (\beta,Q^2,x_{\text{L}})$ with the results from the GBW model (dashed black line) and the bSat model (solid blue line) with the parameters of Fit 1 and 3 respectively.}
	\label{fitplot}
\end{figure*}

\begin{figure*}
	\includegraphics[width=0.45\linewidth]{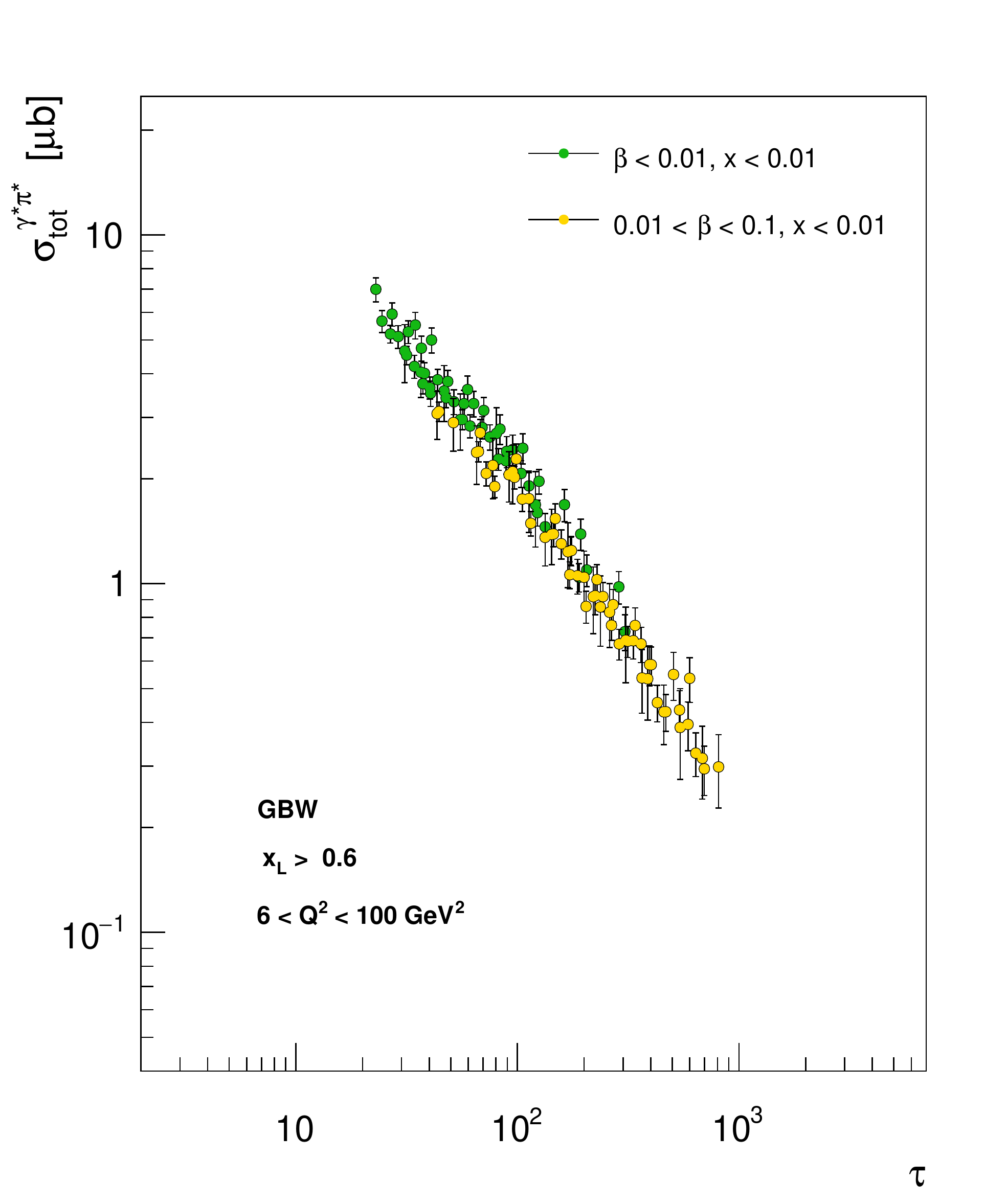}\hskip0.5cm
	\includegraphics[width=0.45\linewidth]{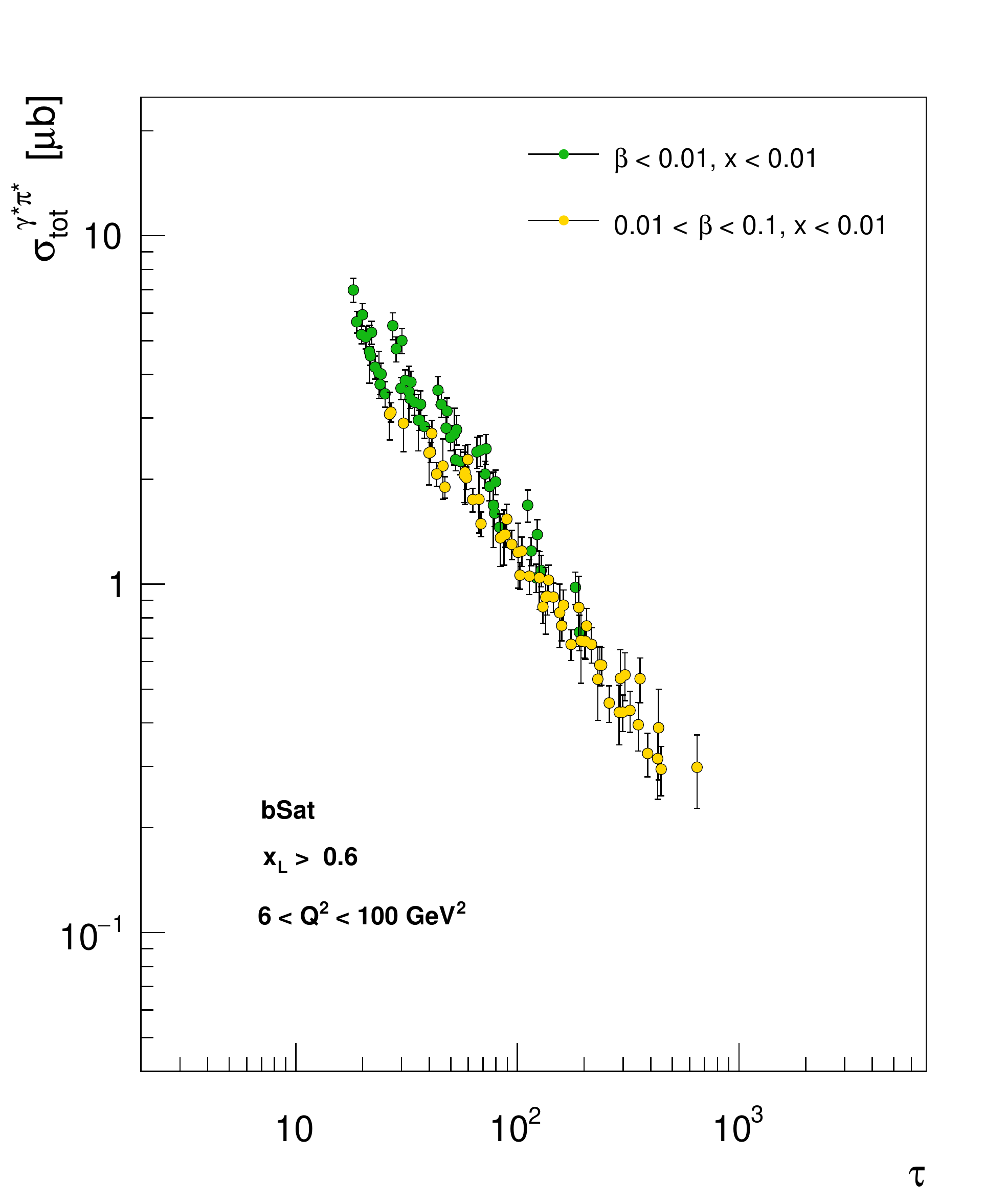}
	\caption{Geometric scaling in experimental data on $\sigma^{\gamma^*\pi^*}$(extracted from $F_2^{LN}$) as a function of $\tau$ from Fit 1 of  the GBW model (left) and Fit  3 of the bSat model (right) in leading neutron events at HERA.}
	\label{expscaling}
\end{figure*}
In Table \ref{table1}, the fit results with the GBW and bSat models are shown. Fit 1 corresponds to the fit where the GBW model  parameters ($\sigma_0,~\lambda,$ and $x_0$) are fitted to the leading neutron structure function data, while for fit 2, these parameters are the same as determined from the usual inclusive DIS data and the assumption of hadronic universality at small $x$ between pion and proton (Eq.\eqref{GBWscaling}) is used and the only parameter is $R_q$, which is determined from the fit. Similarly, fit 3 corresponds to the fit where the gluon density parameters in the bSat model $A_g,~\lambda_g$ and $C$ are fitted to the leading neutron structure function data, while for fit 4, these are the same as determined from the usual inclusive DIS data and the assumption of hadronic universality at small $x$ between pions and protons (Eq.\eqref{bSatscaling}) is used where the only parameter $R_q$ is determined from the fit. The fits are performed using the MINUIT2 package \cite{James:1994vla} and  the corresponding $\chi^2/N_{\text{dof}}$ are shown, where $N_{\text{dof}}$ is $N_{\text{p}}$ - (\# of parameters) and $N_{\text{p}}$ is number of data points in the fit. We see that the GBW model in its original form, in addition to inclusive DIS data, also provide a good description of the leading neutron data for both the scenarios. The bSat model fit having an explicit DGLAP evolution provides the best description of the leading neutron data in all the fits and prefers a slower evolution of the gluon density, as well as almost half the number of gluons as compared to the inclusive DIS case as illustrated in Fit 3. For bSat, fit 4 with hadronic universality assumption also provides a reasonable description of the data.

 In Fig.~\ref{fit_details}, the sensitivity of the fit 1 quality to the choice of the minimal value of the proton’s longitudinal momentum fraction taken by the neutron, $x_{\text{L}}$$_{\text{min}}$, in the data is shown. The blue solid curve represents the variation of $\chi^2/N_{\text{dof}}$ with the cutoff  $x_{\text{L}}$$_{\text{min}}$ keeping the other parameters ($\sigma_0,~\lambda,$ and $x_0$) fixed from fit 1, while the red dashed curve shows the behaviour of  $\chi^2/N_{\text{dof}}$ where all the parameters are kept free in the fit. We see that the choice $ x_{\text{L}}$$_{\text{min}} = 0.6 $ is optimal and the  quality of the fit deteriorates as the $x_{\text{L}}$$_{\text{min}}$ is reduced further. This is not surprising as the one pion exchange approximation(OPE) holds good only for large momentum fractions carried by the neutron.
 
   In Fig.~\ref{satscales}, the saturation scale of the pion, $Q_s^2$, is plotted as a function of Bjorken $x$ at $x_{\text{L}}=0.6$  in the left plot. The saturation scale in the GBW model is extracted from fit 1 as defined in Eq.\eqref{GBWQs} and from fit 3 for bSat model as defined in  Eq.\eqref{bSatQs}. The energy dependence of the saturation scale of the pion is different in both the models, though both the models predict the saturation scale to be  $Q_s^2\sim1$ GeV$^2$ at $x=10^{-6}$. For reference, the ratio of the saturation scale of pions to that of the protons is also shown in the right plot. For the GBW model, we see that the saturation scale of the virtual pion probed in the leading neutron events is about half of the saturation scale of the proton for all $x$ values, and the energy dependence is almost identical for both. For the bSat model, this is not the case  and the saturation scale is half of the value of the proton's saturation scale only at small $x$. The evolution is considerably different for the two models. The small values of saturation scale of pions indicates that it  is less sensitive to saturation and one needs to go to higher energies to probe non-linear effects in leading neutron events as compared to the usual DIS events.
 
 In Fig.~\ref{sigmaGBW}, the normalized dipole cross section ($\sigma/\sigma_0$) in the GBW model with the fit 1 parameters from Table \ref{table1} is plotted, evaluated at different scaled Bjorken variable $\beta = 10^{-2},~10^{-4},~10^{-6}$. In the left plot, the dipole cross section is plotted as a function of dipole size $r$ where we observe that the cross section saturates for large dipole sizes for all values of $\beta$ as expected. In the right plot, the presence of geometric scaling in the GBW model is illustrated as all the curves from left plot with  different $\beta$ values merge into a single curve when the dipole cross section is plotted against the dimensionless variable $rQ_s$ which shows that indeed the dipole cross section is a function of a single dimensionless variable rather than $x$ and $Q^2$ independently. The geometric scaling is exact in the GBW model.

 In Fig.~\ref{sigmabSat}, the dipole cross section $\frac{\dint\sigma^{(\pi)}_{q\bar{q}}}{\dint^2\textbf{b}}$ in the bSat model is plotted, with parameters from the fit 3, evaluated at different values of the scaled Bjorken variable $\beta = 10^{-2},~10^{-4},~10^{-6}$. In the left plot, the dipole cross section is plotted as a function of dipole size $r$, where the cross section saturates for large dipole sizes. For the bSat model, the curves from the left plot with different $\beta$ values do not merge into a single curve when the dipole cross section is plotted against the dimensionless variable $rQ_s$ because of the breaking of geometric scaling. This is primarily due to the explicit DGLAP evolution of the gluon density in the bSat model where the gluon density is evaluated at a scale $\mu^2 = \mu_0^2 +\frac{C}{r^2}$ which depends upon the dipole size $r$,  hence resulting in violation of scaling. The scaling is exact only for large dipole sizes where $\mu^2 \sim \mu_0^2$ as is seen in Fig.~\ref{sigmabSat} for $rQ_s>2.5$. 
 
 In Fig.~\ref{fitplot}, the predictions of the fits 1and 3 of the leading neutron structure function $F_2^{LN}$ as a function of the scaled Bjorken variable,  $\beta$ are  presented, for different values of $x_{\text{L}}$ with varying $Q^2$ in the GBW and bSat dipole models and confronted with the HERA measurements from \cite{H1:2010hym}. Both the models provide a good description of the energy dependence of the data for $x_{\text{L}}\geq 0.6$, while for lower $x_{\text{L}}$ values the models underestimates the data. This is because the models have been fitted for kinematic region $x_{\text{L}}\geq 0.6$, where one pion exchange approximation holds good. The curves for both the GBW and the bSat model are practically on top of each other in the whole kinematic region. 

In Fig.~\ref{expscaling}, the experimental data of the total cross section $\sigma^{\gamma^*\pi^*}$ extracted from the leading neutron structure function, $F_2^{LN}$, employing the one pion exchange approximation is shown for $x_{\text{L}}>0.6$ in the phase space  defined by the photon virtuality in $6\label{key} < Q^2 < 100$ GeV$^2$ and the Bjorken $x$ values in $1.5 \cdot 10^{-4} < x < 3 \cdot 10^{-2}$. The total cross section $\sigma^{\gamma^*\pi^*}$ is plotted against the dimensionless variable $\tau = Q^2/Q_s^2(\beta)$ with saturation scale from Eq.\eqref{GBWQs} in the GBW model in the left plot and from Eq.\eqref{bSatQs} in the bSat model in the right plot . We observe that the experimental data exhibit geometric scaling behavior for the events with $x<0.01,~\beta<0.1$ for the saturation scale values obtained from both GBW and bSat models. For the available data, the total cross section $\sigma^{\gamma^*\pi^*}$ shows the 1/$\tau$ dependence at large $\tau$  which is very similar to what has been discovered for the total cross section  $\sigma^{\gamma^*p}$ in usual DIS events. On the right plot, a few experimental data points at small values of $\tau$ are off from the scaling behaviour. This is due to the different magnitude and the energy dependence of saturation scale in the bSat model as compared to GBW model at moderate $x$ values. The left plot is the main result of this paper where the saturation scale is extracted from the GBW model since geometric scaling is exact in the GBW model while it breaks down in the bSat model where the dipole size affects the evolution of strong coupling constant and gluon density and hence the saturation scale values. It is interesting to note that the scaling region extends to $\beta<0.1$, if we take a closer look at the geometric scaling plot of the inclusive DIS data from \cite{PhysRevLett.86.596}, the experimental data points in the large $x$ region from H1 and ZEUS also show geometric scaling and the data points responsible for the breaking of scaling are from experiments other than HERA. Several other investigations with the analysis of HERA data \cite{Caola:2010cy,Praszalowicz:2012zh} also concluded that geometric scaling  holds for Bjorken $x<0.1$ for inclusive DIS events. The scaling behavior in leading neutron DIS is identical to what has been observed for inclusive DIS events.

\section{CONCLUSIONS AND DISCUSSION}

In  this work, it is shown that the dipole model phenomenology can successfully describe the leading neutron DIS cross section with the new fits of the dipole model to the leading neutron structure function data. Using  OPE, both the GBW and the bSat (IP-Sat) models provide a good description of the leading neutron data. Among all the fits, the fit with bSat model having an explicit DGLAP evolution describes the data best with a $\chi^2/N_{\text{dof}}=1.22$, as depicted in Table \ref{table1}.  The absorptive corrections are not included in this analysis as they are only dominant for small photon virtualities ($Q^2<6$ GeV$^2$) which is outside the kinematic regime considered in this  analysis and would just affect the normalization of the pion flux. In the first part, we have shown the presence of geometric scaling in the GBW model with the new fit parameters of the model from the leading neutron DIS. The scaling behaviour is not exact in the bSat model for small dipole sizes where the evolution effects play a dominant role, while for the large dipole sizes the model shows scaling behaviour similar to the  scaling in the GBW model. The new fit results are used to extract the saturation scale of the pion which in general is about half of the proton's saturation scale and the energy dependence of the pion saturation scale is identical to that of the proton in the GBW model. The main result of this study is presented in left plot of Fig.~\ref{expscaling}, where we have shown, using OPE, that the experimental data on  $\sigma^{\gamma^*\pi^*}$ exhibits geometric scaling over an extended region of $Q^2$ and shows  $1/\tau$ behaviour for large $\tau$ values. The presence of the scaling for $Q^2>Q_s^2$ shows that the geometric scaling is more  general than in the saturation model and can be attributed to the presence of a saturation boundary in the data  which has its root in the evolution equations as described in detail in \cite{IANCU2002327,PhysRevD.66.014013}. This is profoundly similar to what has been previously seen in the $\sigma^{\gamma^*p}$ in inclusive DIS events. The presence of geometric scaling and similar behaviour of $\sigma^{\gamma^*\pi^*}$ as a  function of $\tau$ hints toward the universality of small $x$ structure between pions and protons, though we need more experimental data at smaller $\beta$ values to further validate this statement. Future colliders such as EIC \cite{Accardi:2012qut,AbdulKhalek:2021gbh} and FCC or LHeC \cite{Agostini:2020fmq} have the potential to probe this region of phase space and it will be interesting to see the differences and the similarities between inclusive DIS events and leading neutron DIS.

\section*{ACKNOWLEDGMENTS}
Sincere thanks to Tobias Toll for a careful reading of the manuscript and valuable discussions that helped in a better understanding of this work. This work is supported by the Department of Science \& Technology, India under Grant No. DST/ INSPIRES/03/2018/000344. 

\bibliographystyle{elsarticle-num}
\bibliography{mybibfile}

    \end{document}